\documentclass[reqno,11pt]{amsart}
\usepackage{color}

\usepackage{amssymb,amsthm}
\usepackage[linktocpage]{hyperref}

\usepackage[english]{babel}

\usepackage{graphicx}
\graphicspath{pic/}
\usepackage{psfrag}
\usepackage{graphicx}
\usepackage{caption}
\usepackage{subcaption}

\makeatletter
\def\blfootnote{\xdef\@thefnmark{}\@footnotetext}
\makeatother
\newcommand{\blue}[1]{{\color{blue}{#1}}}

\usepackage{a4}
\numberwithin{equation}{section}

\usepackage[mathscr]{eucal}

\usepackage{version}
\excludeversion{LONG}

\newcounter{mnotecount}[section]
\renewcommand{\themnotecount}{\thesection.\arabic{mnotecount}}
\newcommand{\mnote}[1]{\protect{\stepcounter{mnotecount}}${\raisebox{0.5\baselineskip}[0pt]{\makebox[0pt][c]{\color{magenta}{\tiny\em$\bullet$\themnotecount}}}}$\marginpar{\raggedright\tiny\em$\!\!\!\!\!\!\,\bullet$\themnotecount: #1}\ignorespaces}

\renewcommand{\mnote}[1]{}

\newcommand{\mymnote}[1]{\mnote{\blue{#1}}}
\renewcommand{\mymnote}[1]{}

\renewcommand{\Re}{\mathbb R}
\newcommand{\tr}{\text{tr}}
\newcommand{\half}{\frac{1}{2}}

\newcommand{\eps}{\epsilon}

\newcommand{\Bo}{\mathcal B}
\newcommand{\Sp}{\mathcal S}

\newcommand{\MM}{\mathcal M}

\newcommand{\FF}{\mathcal F}

\newcommand{\Proj}{{\mathbb P}}
\newcommand{\BProj}{\Proj_{\Bo}}

\newcommand{\bodymetric}{b}
\newcommand{\spacemetric}{g}

\newcommand{\norm}{\nu} 
\newcommand{\DD}{\mathcal D} 

\renewcommand{\SS}{\mathcal S}

\theoremstyle{plain}

\newtheorem{remark}{Remark}[section]

\title{Self-gravitating elastic bodies}
\author[L. Andersson]{Lars Andersson} \email{laan@aei.mpg.de}
\address{Albert Einstein Institute, Am M\"uhlenberg 1, D-14476 Potsdam,
  Germany}

\begin{document}
\selectlanguage{english}

\date{July 6, 2014}
%\date{\today \ {\em File:\jobname{.tex}\,}}
%\date{\today \ {\em File:\jobname{.tex} at \currenttime}}

\begin{abstract}
Extended objects in GR are often 
modelled using distributional solutions of the Einstein equations with point-like sources, 
or as  the limit of infinitesimally small ``test'' objects. In this note, I will consider models of finite self-gravitating extended objects, which make it possible to   
give a rigorous treatment of the initial value problem for
 (finite) extended objects.
\end{abstract}

\maketitle

\blfootnote{Based on a talk given at the 2013 WE-Heraeus seminar on ``Equations of motion in relativistic gravity''}

%\tableofcontents

\section{Introduction} \label{sec:intro}
Extended objects in GR are often 
modelled using distributional solutions of the Einstein equations with point-like sources, 
or as  the limit of infinitesimally small ``test'' objects. In this context, 
gravitational
self-force manifests itself through corrections to geodesic motion, in analogy to radiation reaction. 
This is relevant for example in the analysis of extreme mass ration inspirals, see \cite{barack:2009CQGra..26u3001B}. 
See also the papers by Harte \cite{harte:2014arXiv1405.5077H} and Pound \cite{pound:thisvolume} for background on the self-force problem. 
%A survey of the self-force problem can be found in the paper by Harte in this volume, cf. %\cite{harte:2014arXiv1405.5077H}. \mnote{in this volume -- add correct ref} 

A widely studied model for objects with internal structure in general relativity are so-called spinning particles. There are several formal approaches to deriving the corrections to geodesic motion for such object, see \cite{dixon:thisvolume} for a survey. 
%including \cite{dixon:1979igsg.conf..156D}. 
These works rely to a large extent on the study of distributional stress-energy tensors representing the particle-like objects. On the other hand, limiting procedures have been applied to study objects with internal structure by Wald and collaborators, cf.  
\cite{wald:2009arXiv0907.0412W}. 
In this note, I will consider models of finite self-gravitating extended objects, which make it possible to   
give a rigorous treatment of the initial value problem for
 (finite) extended objects. Such models could serve as a basis for the above mentioned limiting considerations. 

%%%%%%%%%%%%%%%%

A serious difficulty in treating self-gravitating material bodies in general relativity, is that matter distributions with finite extent are typically irregular
  at the surface of the body. This phenomenon can be seen already by considering a stationary Newtonian polytrope, with equation of state 
\mnote{add cite -- Binney and Tremaine?} 
$$p = K \rho^{\gamma}
$$
Then the density $\rho$ behaves as 
\mnote{integrate Lane-Emden once or use eq. (3.17) in deformations of compact stars paper} 
$$
\rho(x) \sim 
d^{\frac{1}{\gamma-1}}(x,\partial \Omega)
$$
where $d(x,\partial\Omega)$ is the distance to the boundary of the body. 
Recall that the sound speed $c_s$ for such a polytrope is given by 
\mnote{see \cite[Appendix 1.E]{binney:tremaine:1987gady.book.....B}}
$$
c_s = \sqrt{\frac{dp}{d\rho}} =
\sqrt{K\gamma}\rho^{\frac{\gamma-1}{2}} 
$$ 
and hence $c_s$ tends to zero at $\partial \Omega$. It follows that the hyperbolicity of the Euler equations degenerates at the free boundary, characterized by the vanishing of pressure,  of a typical polytrope in vacuum. In particular the particles at the boundary move as if in free fall. 

In contrast, perfect fluid bodies in vacuum with equation of state such that the density at the free boundary is non-vanishing, are sometimes referred to as liquid bodies. An example of an equation of state of this type is 
$$
p = D(\rho-\rho_0)
$$
where $D, \rho_0$ are suitable constants. For a steady fluid body with this equation of state, the density will be $\rho_0$ at the boundary of the body. In this particular case, we also see that the sound speed does not go to zero at the boundary, and there is no degeneration of hyperbolicity. However, for liquid bodies this is not generally the case. 
See \cite[\S 3.5]{2014arXiv1402.6634A} for discussion.

%%%%%%%%%%%%%%%%%%%%%%%%

For elastic bodies, like liquid bodies, we may expect the density of the material to be non-zero at the boundary, and hence there will be a jump in the density at the surface of the body.
%, i.e. the density will be of the form 
%$\rho \chi_\Omega$, cf. figure \ref{fig:elastbod}. 
Further, for elastic bodies, we may expect that the field equations remain non-degenerate and hyperbolic up to boundary. 
%hyperbolicity does \emph{not} break down break down at the surface 
For elastic bodies, the free boundary condition, which can be formulated as saying that the normal pressure at the boundary vanishes, is known as the zero traction boundary condition. 

%%%%%%%%%%%%%%%%%%%%%%%%%%%
%%%%%%%%%%%%%%%%%%%%%%%%%%%%

%\subsection{Cauchy problem in continuum mechanics}
%We start by discussing the Cauchy problem for infinitely extended materials. 

Following the qualitative discussion above, we shall now mention some results on the Cauchy problem in continuum mechanics. First we consider infinitely extended bodies. For the case of fluids, Christodoulou \cite{christodoulou:shocks} gives conditions for shock formation
  for small data,
while for elastic materials John \cite{john:1984:MR755727} gives condition (genuine
  nonlinearity)  under which
  small data lead to formation of singularities. Sideris \cite{sideris:nullcond}
  gives a version of the null condition for elasticity and proves global existence
  for small data.

For bounded matter distributions, the situation is more complex. 
As mentioned above, for liquid or fluid bodies in vacuum, the  hyperbolicity of the evolution equation  degenerates at boundary. This problem can be overcome by using eg. weighted energy estimates. See   
 \cite{MR2178961,2010arXiv1003.4721C,trakhinin:2008arXiv0810.2612T} for recent work on this problem. The Cauchy problem for elastic bodies with free boundary can in Lagrange coordinates be written as a quasi-linear hyperbolic problem with boundary condition of Neumann type and treated using the methods of eg. \cite{Koch}. See \cite{BWP,2011CQGra..28w5006A}
%andersson:etal:2011arXiv1106.3879A} 
for applications of these techniques in elasticity.

If we on the other hand consider self-gravitating material bodies, much less is known. In fact, apart from some limited results which we shall mention below, the problem of constructing solutions of the initial
  value problem for self-gravitating liquid or fluid bodies in vacuum (both in
  Newtonian gravity and GR) is largely open. 
  
\begin{figure}[!hbt]
\centering
\includegraphics[width=0.48\textwidth]{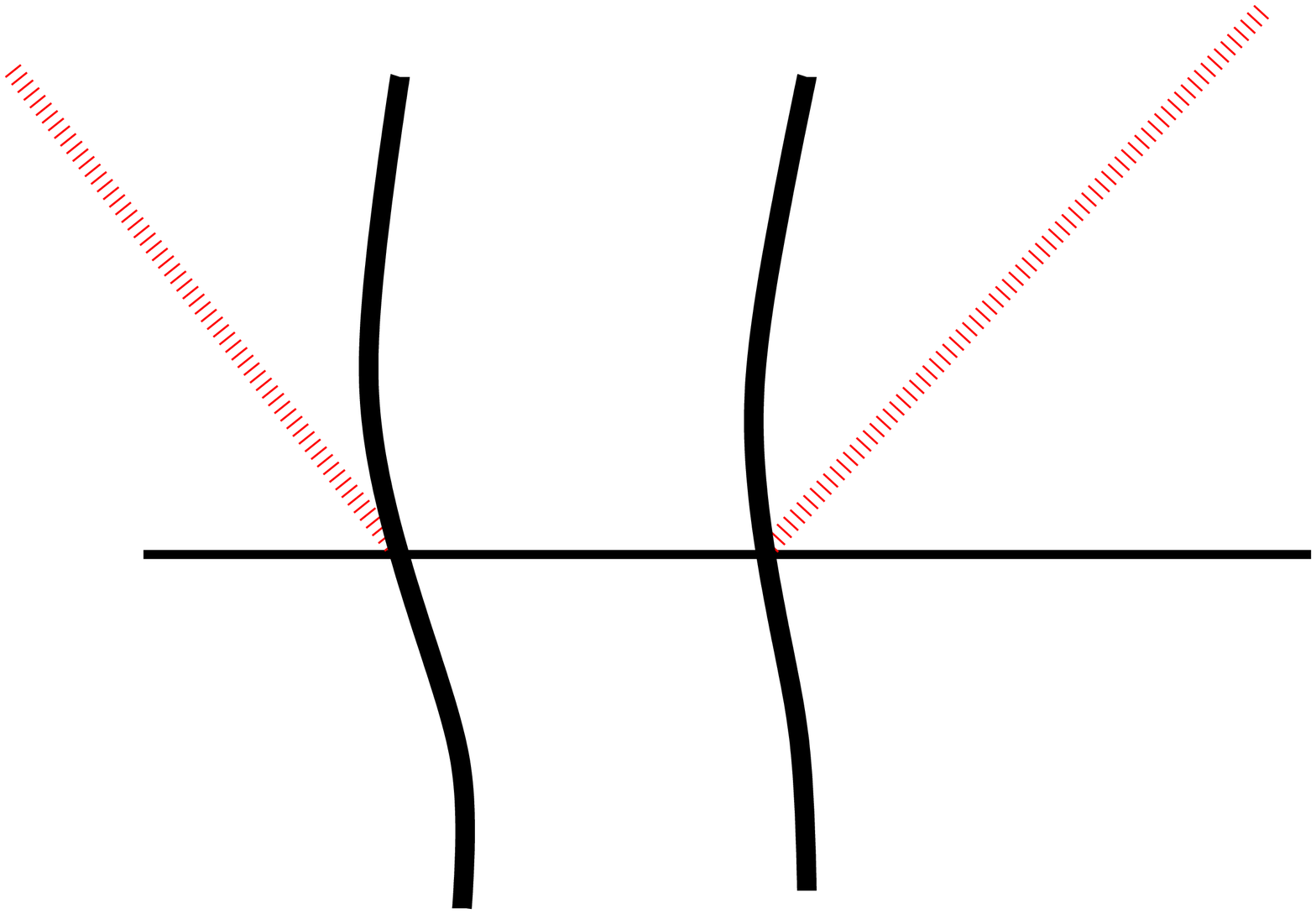}
\caption{}
\label{fig:radiate}
\end{figure}

The Einstein equations imply hyperbolic equations for the components of curvature. Hence the irregularity at the boundary of a self-gravitating body could in general be expected to radiate into the the surrounding spacetime, preventing this from being regular, cf. figure \ref{fig:radiate}. As this clearly does not occur for realistic self-gravitating bodies, there must be a geometric 
``conspiracy'' 
at the boundary of a self-gravitating body undergoing a regular evolution in Einstein gravity. This then has to be reflected in compatibility conditions on the Cauchy data for such a body, see \cite{vanelst_ellis_schmidt_2000PhRvD..62j4023V}. 
%\mnote{mention van Elst, Ellis, Schmidt paper} 
%\mnote{in order to construct dynamical self-gravitating bodies in GR, one must
% understand under which conditions this fails to happen }

It has in recent work been possible to prove local well-posedness for the 
the initial value problem for self-gravitating elastic bodies in Newtonian
  gravity, cf.  \cite{2011CQGra..28w5006A}, and general relativity, cf. \cite{andersson:oliynyk:MR3150755,AOS:dynelast}, see also section \ref{sec:dynelast} below.   
%  
%  For the case of self-gravitating elastic bodies in GR, this is work in progress %\cite{2014arXiv1401.0277A,AOS:dynelast}, see also section \ref{sec:dynelast} below. 
%
In both cases, one finds that  corner conditions on the initial data originating from the  free boundary condition, which from a PDE point of view is of Neumann type, as well as compatibility conditions on the Cauchy data. 

If we turn to dynamical liquid or fluid bodies in general relativity, the results are quite limited. 
Choquet-Bruhat and Friedrich \cite{ChoquetBruhat:2006qn} considered the initial value problem 
for a dust body in Einstein gravity, assuming a  density which is regular at the boundary. The work of Kind and Ehlers \cite{kind:ehlers:1993CQGra..10.2123K} on self-gravitating fluid bodies in general relativity restricts to spherical symmetry but allows a discontinuity at the boundary for the matter density. Rendall \cite{rendall:1992} was able to prove local well-posedness for Einstein-fluid bodies with certain restricted class of equations of state, and with smooth density at the boundary. 

%\subsection{Steady states}
Steady states of self-gravitating bodies provide in particular solutions of the initial value problem, and thus, apart from their intrinsic interest, a study of steady states gives useful information for the study of the dynamics of self-gravitating bodies. 
Steady states of fluid configurations in Newtonian gravity may be
complicated, examples are Dedekind and Jacobi ellipsoids, cf. \cite{chandra:ellipsoidal,meinel:ansorg:book}. 
Lichtenstein \cite{lichtenstein} constructed static and rotating fluid
Newtonian fluid bodies. His results have been extended to elastic matter by Beig and Schmidt 
\cite{beig:schmidt:celest}. For general Newtonian liquid or fluid bodies there are only limited results available. Lindblad and Nordgren proved a priori estimates for incompressible Newtonian fluid bodies \cite{lindblad-2008}. Further, problems of dynamics and stability of self-gravitating fluid and liquid bodies in Newtonian gravity have been studied by
Solonnikov, see eg. \cite{solonnikov:MR2084184,solonnikov:MR2493258} and references therein.

Static self-gravitating fluid bodies are spherically symmetric, in Newtonian gravity as well as in general relativity, cf.  \cite{masood:2007}. Lindblom 
\cite{lindblom:axisym} gave an 
argument showing that viscous stationary 
fluids in GR are axi-symmetric.
Heilig \cite{heilig} 
constructed rotating fluid bodies in GR. 
It is an open problem whether helically symmetric rotating states exist in GR,
  cf. \cite{2007PhRvL..98l1102B,2009GReGr..41.2031B,2009PhRvD..80l4004U} for related work.

Although relativistic elasticity has been studied since shortly after
  the introduction of relativity, cf. \cite{herglotz} (special relativity),
  \cite{rayner, carter:quintana,KM,shadi:elast}, 
until recently no existence or well-posedness results except in the
spherically symmetric case, cf. \cite{park:2000}.
Work by the author with Beig and Schmidt shows that there are examples of static self-gravitating elastic bodies in general relativity which have no symmetries, cf. \cite{ABS}. Similarly, there are rigidly rotating self-gravitating elastic bodies in general relativity with minimal symmetry, i.e. which are stationary and axially symmetric \cite{andersson:beig:schmidt:rotating:MR2583306}.

%\item We see many non-symmetric and ``steady states''
%   around
%  us -- need a consistent model for non-symmetric extended bodies.

%%%%%%%%%%%%%%%%%%%%%%%%%%%%%%%%%

\section{Classical elasticity} \label{sec:classelast} 

An elastic body
is described in terms of configurations with respect to 
a \emph{reference body} $\Bo$, a domain in the extended body $\Re^3_{\Bo}$.

\begin{figure}[!ht]
\psfrag{MAB}{$\Bo$}
\psfrag{MAfinvB}{$f^{-1}(\Bo)$}
\psfrag{MAR3B}{$\Re^3_{\Bo}, x^A$}
\psfrag{MAR3S}{$\MM, x^\mu = (t, x^i), g_{\mu\nu}$}
\psfrag{MAf}{$f^A$} 
\psfrag{MAphi}{$\phi^\mu$}

\includegraphics[width=0.9\textwidth,height=3in]{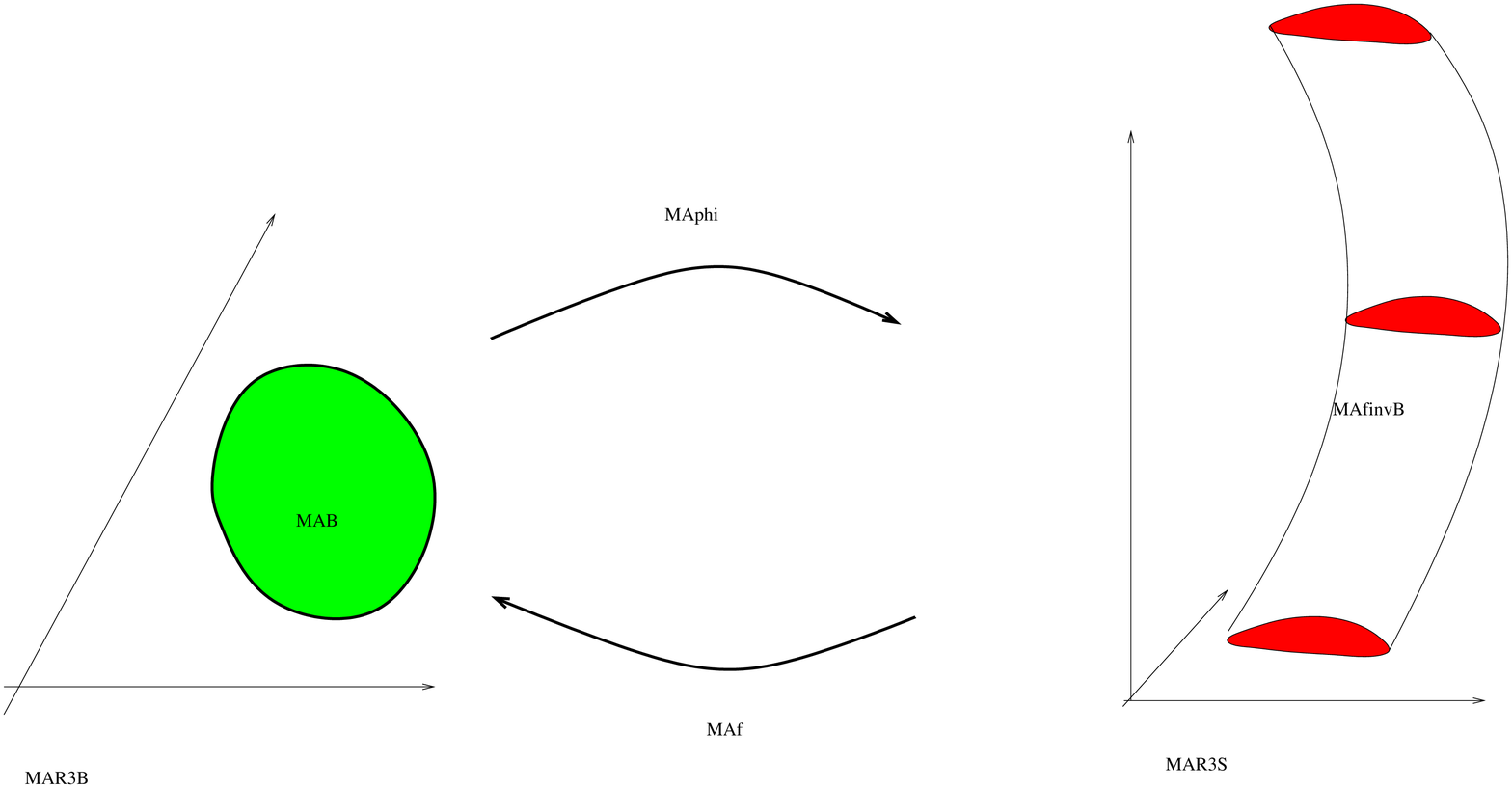} 
\end{figure} 

The \emph{configuration maps} $f$ from the physical spacetime to the reference body, and the \emph{deformation maps} $\phi$ from the reference body to spacetime are assumed to satisfy 
$$
f \circ \phi \big{|}_{\Bo} = \mathbf{id} .
$$
The role of the configuration map $f$ in the Eulerian variational formulation of elasticity in the context of general relativity has been stressed by Kijowski and Magli \cite{KM}. See the books by Marsden and Hughes \cite{hughes:marsden} and Truesdell and Noll \cite{truesdell:noll:MR2056350} for background on elasticity. 
 
The physical body $f^{-1}(\Bo)$ moves in spacetime $\MM$ with coordinates 
$x^\mu = (t, x^i)$ and metric  $g_{\mu\nu}$. Coordinates as well as coordinate indices on $\Bo$ are denoted with capital letters, $X^A$. It is convenient to endow the body $\Bo$ with a \emph{body metric} $\bodymetric_{AB}$. For many situations, this can be taken to be the Euclidean metric $\bodymetric_{AB} = \delta_{AB}$.

%\item space (or space-time): $\MM$ = $\Re^3_{\Sp}$ (or $\MM = \Re^3_{\Sp} \times
%  \Re_{\text{time}}$), coordinates $x^\mu = (t,x^i)$.  

We start by considering the non-relativistic case. In the non-relativistic case it is natural to take $\MM = \Re_t \times \Re^3_\Sp$, where $\Re^3_\Sp$ is the space-manifold, metric $\spacemetric_{ij}$, which in the non-relativistic case can be taken to be Euclidean. 
The action for a hyperelastic body in Newtonian gravity takes the form 
\begin{equation}\label{eq:hyperelastic}
S = \int \Lambda dt d^3 x 
\end{equation} 
where 
\begin{equation}\label{eq:Lambda} 
\Lambda = \Lambda^{kin} - [ \Lambda^{grav} + \Lambda^{pot} + \Lambda^{elast}
]  
\end{equation} 
where 
\begin{align*} 
\Lambda^{kin} &= \half \rho v^2 \chi_{f^{-1}(\Bo)} , \\ 
\Lambda^{grav} &= \frac{|\nabla U|^2}{8\pi G} ,\\
\Lambda^{pot} &= \rho  U \chi_{f^{-1}(\Bo)},\\ 
\Lambda^{elast} &=  n \epsilon \chi_{f^{-1}(\Bo)} \,.
\end{align*} 
See \cite[\S 3]{2014arXiv1402.6634A}.
Here $n = \det \partial f$ is the number density, and 
$\eps = \eps(f, \partial f)$ is the stored energy function, representing the internal energy of the material. We have, for clarity included the indicator function $\chi_{f^{-1}(\Bo)} = \chi_\Bo \circ f$ of the physical body, where $\chi(X) = 1$ for $X \in \Bo$, and $\chi(X) = 0$ otherwise. 
The physical mass density is  
$\rho = nm$ where $m$ is the specific mass of the material particles.
Further, $U$ is the Newtonian potential and 
$$
|\nabla U|^2 = \partial_i U \partial_j U \spacemetric^{ij} .
$$
 The kinetic term in the action is 
defined in terms of the square velocity 
$$
v^2 = v^i v^j \spacemetric_{ij} ,
$$
with the 3-velocity, given by $v^i = - \phi^i{}_{,A} f^A{}_{,t}$, representing the motion in space of the material particles.
It should be stressed that the terms $\Lambda^{pot}, \Lambda^{kin},
\Lambda^{elast}$ are supported on $f^{-1}(\Bo)$ while the term
$\Lambda^{grav}$ should be viewed having support on the whole $\Re_{\Sp}$. 

\begin{remark} 
\begin{enumerate} 
\item Defining the Newtonian potential by the Poisson integral 
\begin{equation}\label{eq:poisson-int}
U(x) = -G\int_{f^{-1}(B)} \frac{\rho(x')}{|x-x'|} d^3 x' ,
\end{equation} 
the term $\Lambda^{grav}+\Lambda^{pot}$ can be replaced by 
$$
\half \rho U \chi_{f^{-1}(\Bo)} 
$$
\item The Lagrangian given in \eqref{eq:Lambda} is of the familiar form 
$$
L = T - V
$$
with $T, V$ the kinetic and potential terms, respectively. The corresponding Hamiltonian (or energy) is then 
$$
H = T + V
$$
%The expression for total energy given in \cite[Appendix
%  5.B]{B&T:1987gady.book.....B} is the Hamiltonian.
\end{enumerate} 
\end{remark} 
The elastic stress tensor is 
$$
%\tau_{AB} = 2 \frac{\partial \eps}{\partial \gamma^{AB}}, \quad \tau_{ij} = n
%\tau_{AB} f^A{}_{,i} f^B{}_{,j} 
\tau_j{}^i = n \frac{\partial \eps}{\partial f^A{}_{,i}} f^A{}_{,j}
$$
This is the canonical energy-momentum tensor for the elastic part of the action. 
Assuming suitable asymptotic behavior for the fields, the Euler-Lagrange equation for the action \eqref{eq:hyperelastic} is 
\begin{equation}\label{eq:E-L-hyperelastic-chi}
\rho v^\mu \partial_\mu v_i \chi_{f^{-1}(\Bo)} + \partial_j (\tau_i{}^j \chi_{f^{-1}(\Bo)} + 
\rho \partial_i U \chi_{f^{-1}(\Bo)} = 0 
\end{equation} 
Now, an important fact is that the divergence 
$$
\partial_j (\tau_i{}^j \chi_{f^{-1}(\Bo)} )
$$
is a function in $L^p$ only if the normal stress vanishes at the boundary of the body, i.e.  
$$
\tau_i{}^j n_j \big{|}_{\partial f^{-1}(\Bo)} = 0, 
$$
cf. \cite[Lemma 2.2]{ABS}. This is due to the fact that the gradient of the indicator function is of the form 
$$
\partial_i \chi^{f^{-1}(\Bo)} = - \norm_i \delta_{\partial f^{-1}(\Bo)} 
$$
where $\delta_{\partial f^{-1}(\Bo)}$ is the surface delta function. 
Thus, 
\begin{subequations}\label{eq:E-L-hyperelastic}
\begin{align} 
\rho v^\mu \partial_\mu v_i + \partial_j \tau_i{}^j  + \rho \partial_i U &= 0 , \qquad \text{ in $f^{-1}(\Bo)$}, \label{eq:hyperelastic-PDE} \\
\tau_i{}^j n_j \big{|}_{\partial f^{-1}(\Bo)} &= 0 \qquad \text{ on $\partial f^{-1}(\Bo)$} \label{eq:hyperelastic-free-boundary} \\
\intertext{coupled to the Poisson equation} 
\Delta U &= 4\pi G \rho \chi_{f^{-1}(\Bo)} \label{eq:poisson-eq}
\end{align}
which has solution given by \eqref{eq:poisson-int}. 
\end{subequations}  
Here 
$$
v^\mu \partial_\mu = \partial_t + v^i \partial_i
$$ 
so that 
$$
v^\mu \partial_\mu v^i
$$ 
gives the acceleration of the physical particles. Equation \eqref{eq:hyperelastic-PDE} corresponds to Newton's force law $F =
  ma$, where now the force includes both force generated by elastic stress as well as the gravitational force, together with the free boundary, or zero traction, boundary condition \eqref{eq:hyperelastic-free-boundary}. The boundary condition represents the fact that the motion of the boundary is not subject to any external forces.  

\mnote{discuss relation to fluids} 

We recall some facts from potential theory. We can write the Newtonian (volume) potential given by \eqref{eq:poisson-int} as  
$$
U = \Delta^{-1} (4\pi G \rho \chi_{f^{-1}(\Bo)})
$$
Differentiating gives  
\begin{equation} \label{eq:dVV-euc} 
\partial_{x^i} U =  \Delta^{-1}[\partial_{x^i} 4\pi G \rho] - \SS[
  \tr_{\partial f^{-1}(\Bo)} 4\pi G \rho
  \norm^i ] ,
\end{equation} 
where $\SS$ is the layer potential and $\norm^i$ is the normal to $\partial f^{-1}(\Bo)$. 
%$$
%\SS [f] (x) = - \frac{1}{4\pi} \int_{\dO} \frac{1}{|x-y|} f(y) d\sigma(y), \quad x \notin \dO
%$$
Similarly, $\partial_{x^i} \SS$ can be expressed in terms of the double
layer potential $\DD$. 
Standard estimates for $\SS, \DD$ and an inductive
argument can be used to estimate $U$. 
%gives 
%$$
%||U||_{W^{k+2,p}(f^{-1}(\Bo))} \leq C ||\rho||_{W^{k,p}(f^{-1}(\Bo))} 
%$$
Due to the jump in the matter density 
$
\rho \chi_{f^{-1}}(\Bo)
$
we have that $\partial^2 U$ is discontinuous at $\partial f^{-1}
(\Bo)$. 
However, $U$ has full regularity up to $\partial
  f^{-1}(\Bo)$. See \cite[Appendix A]{2014arXiv1402.6634A} for details.

In the material frame (Lagrange coordinates) the physical body is represented by the deformation map $\phi(\Bo)$. The material form of the action is got by simply pulling back the Lagrange density from the Eulerian picture (in spacetime) to get 
$$
S_{material} = \int \phi^* (\Lambda dt d^3x).  
$$
The Euler-Lagrange equation can then be calculated purely in the material picture. An important simplification is gained due to the fact that the domain of the body in the material picture is the reference body $\Bo$, which is time-independent. One finds that under suitable assumptions on the stored energy function, the Cauchy problem for the elastic body in material frame is  
an initial-boundary value problem on $\Bo$ with Neumann
  type boundary conditions. 
%The initial data must satisfy compatibility
%(corner)  conditions at $\partial \Bo$. 

Since we have $\tau_i{}^j = \tau_i{}^j(f, \partial f)$, the expression $\partial_j \tau_i{}^j$ is a quasi-linear second order operator on $f$. 
Disregarding the gravitational self-interaction for the moment, hyperbolicity of the system
\eqref{eq:E-L-hyperelastic}
is determined by the properties of the \emph{elasticity tensor} 
$$
L_A{}^i{}_B{}^j = \frac{\partial^2 \eps}{\partial f^A{}_{,i} \partial
  f^B{}_{,j}}
$$
eg. 
rank-one positivity
$$
L_A{}^i{}_B{}^j \xi^A\eta_i \xi^B \eta_j \geq C |\xi|^2 |\eta|^2 
$$ 
or 
pointwise stability 
$$
L_A{}^i{}_B{}^j \xi^A{}_i \xi^B{}_j \geq C 
\xi^A{}_i \xi^B{}_j \bodymetric_{AB} \spacemetric^{ij} 
%|\bar \xi|^2, \quad \text{ where } 
%\bar \xi^{AB} = \xi^{(A}{}_i f^{B)}{}_{,j} \delta^{ij}  
$$
where $C$ is some positive constant.
If one of these conditions hold, the system \eqref{eq:E-L-hyperelastic} forms a quasi-linear elliptic-hyperbolic system with Neumann-type boundary conditions. 

A formulation of elasticity compatible with general relativity requires the elastic action to be generally covariant. This implies that the stored energy function is \emph{frame indifferent}. Define the strain tensor $\gamma^{AB}$ by 
$$
\gamma^{AB} = f^A{}_{,i} f^B{}_{,j} \spacemetric^{ij}
$$
and let $\lambda_i$,
$i=1,2,3$ be the fundamental invariants of 
$\gamma^A{}_B = \gamma^{AC} (\bodymetric)_{CB}$. 
The material is frame indifferent if $\eps = \eps(f, \gamma^{AB})$ and 
isotropic if $\eps = \eps(\lambda_i)$

\begin{remark}
\begin{enumerate} 
\item In the variational problem of classical elasticity (with energy determined purely by the elastic term), polyconvexity \cite{ball:1977}, i.e. the condition 
$$
\eps(F) = \hat \eps(F,\text{Cof} F, \det F)
$$
where $F = (\phi^{i}_{,A})$, with $\hat \eps$ convex, leads to cancellations which in certain circumstances allow one to show convergence of minimizing sequences. 
\item Small perturbations around a stress free state are governed by the quasi-linear wave
  equation 
\newcommand{\grad}{\text{grad}}
\renewcommand{\div}{\text{div}}
$$
\partial_t^2 \phi - c_2^2 \Delta \phi - (c_1^2 - c_2^2) \grad \, \div \phi =
F(\nabla \phi, \nabla^2 \phi), 
$$
cf. \cite{agemi:2000InMat.142..225A}.
\mnote{\cite{agemi:2000InMat.142..225A} mentions John and Shatah showed that if no plane wave is
genuinely nonlinear in the sense of John, then null condition holds}
\item The field equation of classical elasiticity is analogous to membrane equation which has 
action 
$$
S = \int \sqrt{|\phi^* g|} 
$$
For the vacuum Einstein equation in wave coordinates, $L^2$ bounded curvature
  (which corresponds to $H^2$ regular data) implies 
  local well-posedness \cite{klainerman:etal:2012arXiv1204.1767K}. 
  For elasticity and the membrane equation, the analogous result would be
  well-posedness for $H^3$ regular initial data. 
  \mnote{Ettinger has announced results on the membrane equation} 
\end{enumerate} 
\end{remark}

A static body is in equilibrium, in particular, the elastic load must balance the load from eg. the gravitational force. 
Further, in Newtonian gravity,  Newton’s principle \emph{actio est reactio} implies further that each component of a body must be in equibrium. The following equilibration condition is a consequence of the assumption that the total load on a body from elastic stress and gravitational force does not generate a motion.  
Gauss' law and the zero traction bundary condition gives for any
  Euclidean Killing field $\xi^i$ with $\xi_{i,j} = \xi_{[i,j]}$
$$
\int_{f^{-1}(\Bo)} \xi^j \partial_i \tau_j{}^i = \int_{\partial f^{-1}(\Bo)}
\xi^j \tau_j{}^i n_i = 0, 
$$
The body is static if the stress load balances the gravitational load 
$$
\partial_i \tau_j{}^i = b_i := \rho \partial_j U
$$
In particular such a load must be \emph{equilibrated} 
$$
\int_{f^{-1}(\Bo)} \xi^i b_i = 0, 
$$
for any Killing field $\xi^i$. 
For a general load this is a non-trivial condition, but a gravitational load is automatically equilibrated. 

%%%%%%%%%%%%%%%%%%%%%%%%%%%%%%%%%%%%%%%%%%%%%%%%%%%%%%%%%%%%%%%%%%%%%%%%%%%%%

%%%%%%%%%%%%%%%%%%%%%%%%%%%%%%%%%%%%%%%%%%%%%%%%%%%%%%%%%%%%%%%%%%%%%%%%%%%%%

As mentioned above, it is convenient in applying PDE techniques to elastic bodies, to consider the system in the material frame. This is true both in the construction of steady states of Newtonian elasticity, see \cite{beig:schmidt:celest} and references therein,  but also for the Cauchy problem. Assuming suitable constitutive relations, the 
initial value problem for a Newtonian self-gravitating body in material frame
is an
elliptic-hyperbolic system with Neumann type boundary conditions.
Well-posedness has been proved for this system in \cite{2011CQGra..28w5006A}, assuming suitable constitutive relations. This result gives the first
  construction of self-gravitating dynamical extended bodies with 
\emph{no symmetries.}
One finds that the initial data must satisfy \emph{compatibility conditions} induced
  by the Neumann boundary conditions. 

%%%%%%%%%%%%%%%%%%%%%%%%%%%%%%%%%%%%%%%%%%%%%%%%%%%%%%%%%%%%%%%%%%%%%%%%%%%%%

\section{Elastic bodies in general relativity} 
The action for an general relativistic elastic body is 
\begin{equation}\label{eq:action-einstein-elastic}
S =  - \int \frac{R \sqrt{-g}}{16\pi G} d^4 x + \int \Lambda \sqrt{-g} d^4 x \,.
\end{equation} 
where $\Lambda = \Lambda(f, \partial f, g) = n \eps \chi_{f^{-1}(\Bo)}$ 
is the energy density of the
material in its own rest frame.  Here we have included the indicator function $\chi_{f^{-1}(\Bo)}$ for space-time trajectory of the body explicitely in the action.  The relativistic number density is given by  
$n = \det(\gamma^{AB})^{1/2}$ with $\gamma^{AB} = f^A{}_{,\mu}
  f^B{}_{,\nu} g^{\mu\nu}$, and 
$\eps$ is the stored energy function. 
As mentioned above, general covariance demands \emph{frame invariance}, i.e. 
$\eps = \eps(f, \gamma^{AB}, g) $.

The  Euler-Lagrange equations for this action are the 
Einstein equations
\begin{subequations}\label{eq:einstein-system} 
\begin{equation}\label{eq:Einstein-elastic}
G_{\mu\nu} = 8 \pi G T_{\mu\nu} \chi_{f^{-1}(\Bo)} \,,
\end{equation} 
where 
\begin{align*}
G_{\mu\nu} &= R_{\mu\nu} - \half R g_{\mu\nu}, \quad
T_{\mu\nu} = 2\frac{\partial \Lambda}{\partial g^{\mu\nu}} - \Lambda
  g_{\mu\nu} 
  %\,. \quad \Rightarrow \nabla^\mu T_{\mu\nu} = 0
\end{align*}
The elasticity equations, including the free boundary condition
$$
T_{\mu\nu} \norm^\nu \big{|}_{\partial f^{-1}(\Bo)} = 0,
$$ 
where $\norm^\nu$ is the normal to the (typically time-like) boundary of the spacetime domain of the body 
are consequences of the conservation equation
\begin{equation}\label{eq:conservation} 
\nabla^\mu (T_{\mu\nu} \chi_{f^{-1}(\Bo)} = 0
\end{equation}
\end{subequations} 
which in turn follows from the Einstein equation \eqref{eq:Einstein-elastic}, but which can also be derived as the Euler-Lagrange equation for the action with respect to variations of the configuration map. The field equations for a general relativistic elastic body may thus be viewed as the Einstein equation \eqref{eq:Einstein-elastic} or, equivalently, as the coupled system \eqref{eq:einstein-system}.

%$$
%
%$$

%\item Let $\gamma^{AB} = f^A{}_{,\mu} f^B{}_{,\nu} g^{\mu\nu}$. The
%  relativistic number density is $n = \det(\gamma^{AB})^{1/2}$, and the
%  stored energy function $\eps$ is given by 
%$\Lambda = n \eps$. 

%%%%%%%%%%%%%%%%%%%%%%%%%%%%%%%%%%%%%%%%%%%%%%%%%%%%%%%%%%%%%%%%%%%%%%%%%%%%%

%%%%%%%%%%%%%%%%%%%%%%%%%%%%%%%%%%%%%%%%%%%%%%%%%%%%%%%%%%%%%%%%%%%%%%%%%%%%%
\subsection{Static body in GR}
We next consider the case of static self-gravitating bodies in general relativity. Thus, we 
assume $(\MM,g_{\mu\nu})$ is static, i.e. there is a global timelike, hypersurface orthogonal, Killing field $\xi^\mu$. Then we have that $\MM = \Re\times
  M$ and we may introduce coordinates $x^\mu = (t, x^i)$ such that the Killing field ix $\xi^\mu \partial_\mu = \partial_t$, with norm $e^{2U} =
  - \xi^\mu \xi_\mu$. For a static spacetime we can write 
\begin{equation*}%\label{eq:ds2}
g_{\alpha\beta} dx^\alpha dx^\beta = - e^{2 U} dt^2 + e^{-2  U}
h_{ij} dx^i dx^j
\end{equation*}
where $U, h_{ij}$ depend only on $x^i$. Kaluza-Klein reduction applied to \eqref{eq:action-einstein-elastic} gives the
action 
\begin{equation}\label{eq:Lstat}
S =  - \int_M \frac{1}{16\pi G}\sqrt{h} (R_h - 2 |\nabla U|_h^2) + \int_M e^{
  U} n \eps \sqrt{h}
\end{equation}
 
%%%%%%%%%%%%%%%%%%%%%%%%%%%%%%%%%%%%%%%%%%%%%%%%%%%%%%%%%%%%%%%%%%%%%%%%%%%%%

%%%%%%%%%%%%%%%%%%%%%%%%%%%%%%%%%%%%%%%%%%%%%%%%%%%%%%%%%%%%%%%%%%%%%%%%%%%%%
The Euler-Lagrange equations are 
%\begin{subequations} \label{eq:Euler}
\begin{align*}
\nabla_j (e^U \sigma_i{}^j) &= e^U (n \epsilon - \sigma_l{}^l) \nabla_i U
\quad \text{\rm in } f^{-1}(\Bo), \quad \sigma_i{}^j n_j |_{f^{-1}(\partial
\Bo)}=0 
%\label{elast} 
\\ 
\Delta_h U &= 4 \pi G e^U(n \epsilon - \sigma_l{}^l)
\chi_{f^{-1}(\mathcal{B})}\quad \text{\rm in } \Re^3_{\Sp}
%\label{potential}
 \\
G_{ij} &= 8 \pi G ( \Theta_{ij} - e^U
\sigma_{ij}\,\chi_{f^{-1}(\mathcal{B})})\quad \text{\rm in } \Re^3_{\Sp}
%\label{metric}
\end{align*}
%\end{subequations}
where
\begin{equation*}% \label{theta}
\Theta_{ij} = \frac{1}{4 \pi G}[\nabla_i U \nabla_j U - \frac{1}{2} h_{ij}
|\nabla U|^2].
\end{equation*}
This system is equivalent to the 3+1 dimensional Einstein equations for
  the static elastic body. 
%%%%%%%%%%%%%%%%%%%%%%%%%%%%%%%%%%%%%%%%%%%%%%%%%%%%%%%%%%%%%%%%%%%%%%%%%%%%%

%%%%%%%%%%%%%%%%%%%%%%%%%%%%%%%%%%%%%%%%%%%%%%%%%%%%%%%%%%%%%%%%%%%%%%%%%%%%%
Let a relaxed reference body $\Bo$ be given. For small $G$, we construct a
  static self-gravitating body, i.e. a solution to the static
  Einstein-elastic equations, which is a deformation of $\Bo$,
  cf. \cite{ABS}.
The construction is carried out in the material frame. Working in harmonic coordinate gauge, the reduced Einstein-elastic system can be cast in the form 
$$
\FF(G,Z) = 0,
$$
where $G$ is Newton’s constant and $Z$ denotes the fields in the material frame version of the system, i.e. the deformation map $\phi$ as well as the material version of the Newtonian potential $U$ and the 3-metric $h_{ij}$. Assuming suitable constitutive relations, the reduced system of
Einstein-elastic equations is an elliptic boundary value problem with Neumann type boundary condition.
Given a relaxed background configuration $Z_0$, which can be viewed as a solution of the Einstein-elastic system with Newtons constant $G = 0$, we would like to apply the implicit function theorem to construct solutions to \eqref{eq:einstein-system} for small $G$.  

However, an obstacle to doing so is the fact that the linearized operator $D_Z \FF(0, Z_0)$ necessarily fails to be an isomorphism. In fact, due to invariance properties of the the equilibration condition, the infinitesimal Euclidean motions, i.e. the Killing vector fields on Euclidean 3-space, are in the kernel. Further, due to the linearized operator $D_Z \FF(0,Z_0)$ has a non-trivial co-kernel, which also corresponds to the infinitesimal Euclidean motions. This is due to the fact that the 
the linearized elasticity operator at the reference configuration is
  automatically equilibrated. Thus, we have a kernel and cokernel
  corresponding to the Killing fields of the Euclidean
  reference metric on $M$ and on $\Re^3_{\Bo}$. 
Applying a projection to $D_Z \FF(0,Z_0)$ in order to get an isomorphim we are in a position to apply the  implicit function theorem to construct a solution for small $G$ to the
  projected system  
\begin{equation*}%\label{eq:redproj}
\BProj \FF(G, Z) = 0 .
\end{equation*}
The proof is completed by showing 
that the solution to the projected system is 
\emph{automatically equilibrated}, i.e. it is a solution to the full system, including the harmonic coordinate condition.

By choosing the reference body to be non-symmetric, we thus get the first construction of self-gravitating static elastic bodies in general relativity \emph{with no symmetries}. Outside the body, the spacetime is a solution of the vacuum Einstein equations, which will be asymptotically flat, but with no Killing vector fields except for the static Killing field.  

%\item Main step in the proof is an analysis of the tension field 
%$$
%V^i = h^{mn} \Gamma^i_{mn}
%$$
%%%%%%%%%%%%%%%%%%%%%%%%%%%%%%%%%%%%%%%%%%%%%%%%%%%%%%%%%%%%%%%%%%%%%%%%%%%%%

\begin{figure}[!hbt]
\centering
\begin{subfigure}{0.45\textwidth}
\includegraphics[width=.8\textwidth]{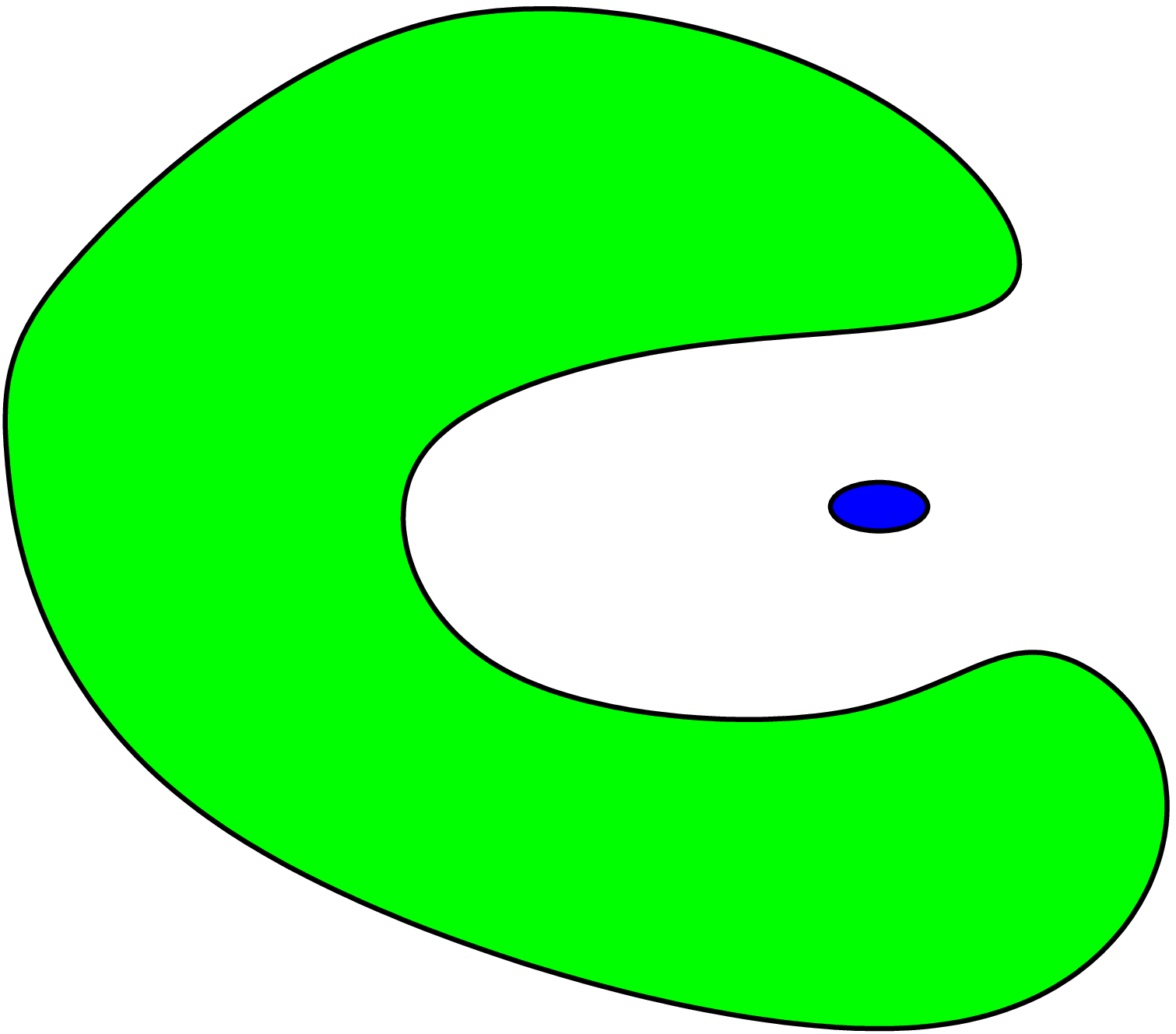}
%\caption{}
%\label{fig:small_outside} 
\end{subfigure} 
\quad 
\begin{subfigure}{0.45\textwidth} 
\includegraphics[width=.8\textwidth]{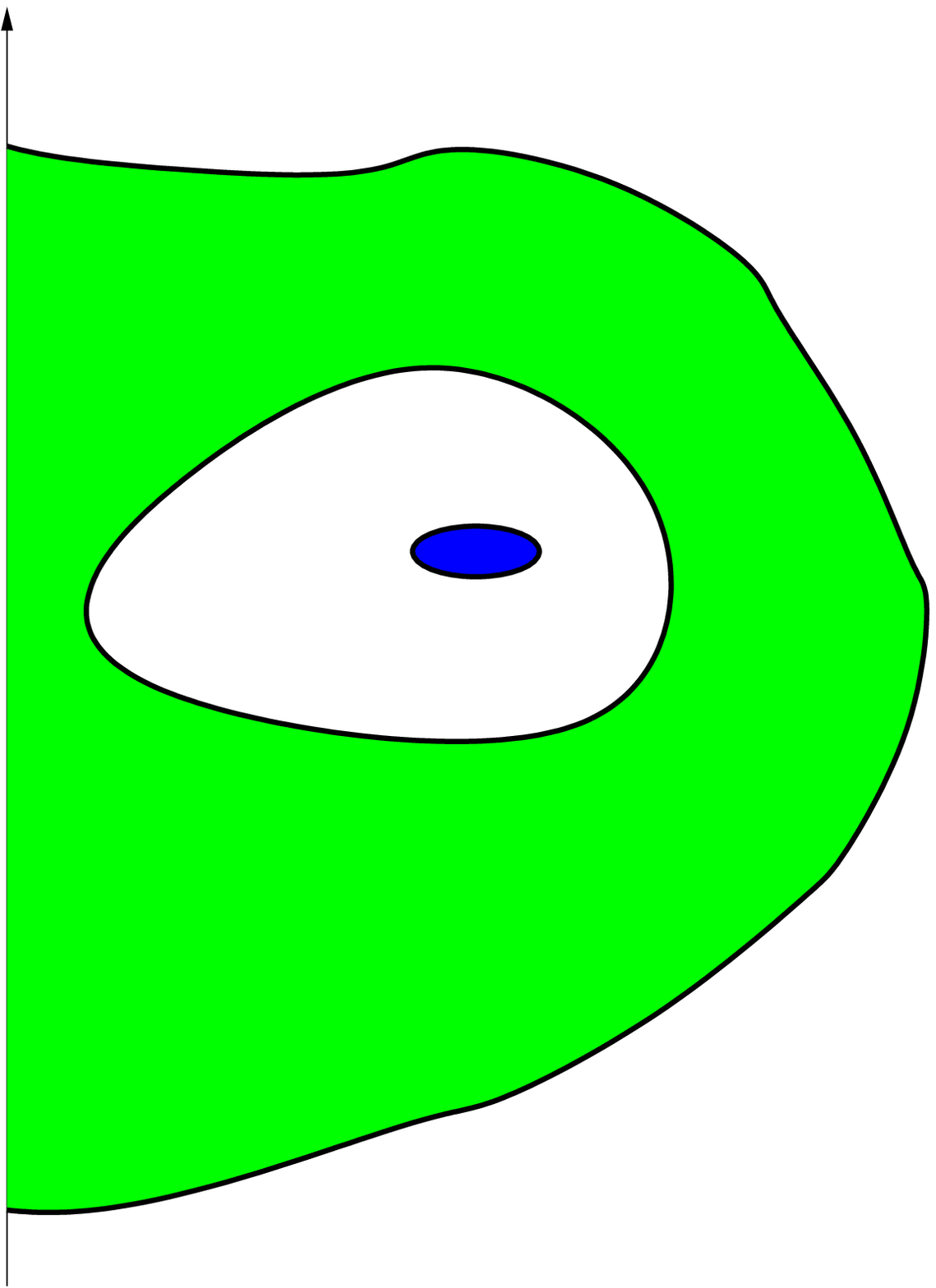}
\end{subfigure} 
\caption{Examples of two-body configurations in equilibrium.}
\label{fig:2-body}
\end{figure}

In Newtonian gravity there are many examples of static self-gravitating many-body systems, consisting of rigid bodies of the type shown schematically in figure \ref{fig:2-body}.
The method described above in the case of static self-gravitatating bodies extends to 
$N$-body configurations \cite{2009CQGra..26p5007A}. In this case, one takes a Newtonian static configuration $N$-body configuration consisting of rigid, self-gravitating bodies as the starting point. Under some conditions on the Newtonian potential one can apply a deformation technique related to that used in the construction of static self-gravitating bodies to construct $N$-body configurations. A particular case consist of placing a small body at a stationary point of the gravitational potential of a large body.

\begin{figure}[!hbt]
%\psfrag{MArho}[cc][cc]{$\rho$} 
 \centering
\includegraphics[width=0.6\textwidth]{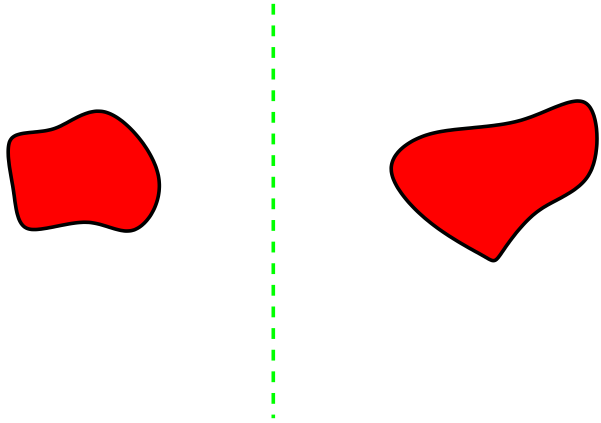}
\caption{Bodies separated by a plane cannot be in equilibrium in Newtonian gravity. This holds in GR if the plane is replaced by a totally geodesic hypersurface. It would be interesting to find a more general characterization of what static configurations are possible.}
\label{fig:separate}
\end{figure}

The proof makes use of the additional degree of freedom corresponding to
  the difference in the centers of mass and alignments of the bodies to
  achieve equilibration. 
In Newtonian gravity, one proves easily that a two bodies separated by
  a plane cannot be in static equilibrium, cf. figure \ref{fig:separate}. 
This relates to Newton's principle
  \emph{actio est reactio}, also mentioned above, which implies that each body must be equilibrated
  with respect to its own self-gravity.

In general relativity, we lack the concept of force (see however 
\cite{cederbaum:2012arXiv1210.4436C} 
for related ideas in the static case)
%-- define ``force'' in static case using
%harmonic coordinates) 
and the problem of characterizing
``allowed''  $n$-body configurations is open. 
Partial results on this problem have been proved by Beig and Schoen \cite{2009CQGra..26g5014B}, and Beig, Gibbons and Schoen  
\cite{2009CQGra..26v5013B}. In particular, bodies separated by a totally geodesic surface
  cannot be in static equilibrium. 
%-- replace by 
%``separating plane in harmonic coordinates''?. 

%%%%%%%%%%%%%%%%%%%%%%%%%%%%%%%%%%%%%%%%%%%%%%%%%%%%%%%%%%%%%%%%%%%%%%%%%%%%%

In order to describe rotating, self-gravitating bodies, we must consider stationary spacetimes, i.e. spacetimes with a Killing field which in the relevant situation will be timelike, but not hypersurface orthogonal. In this case, 
Kaluza-Klein reduction gives action  
\begin{align*}
S = - \int_M \frac{\sqrt{h}}{16\pi G} \left (R_h  - 2 |D U|_h^2
+ e^{4 U}  |\omega|_h^2 \right ) + \int_M n\eps \, e^U \sqrt{h} \,,
\end{align*}
In this case, one may use techniques related to those discussed above to construct self-gravitating rotating bodies in general relativity as deformations of axi-symmetric relaxed, non-rotating, 
reference states, see \cite{andersson:beig:schmidt:rotating:MR2583306}. By choosing the reference body appropriately we get rigidly rotating self-gravitating elastic bodies with a minimal amount of symmetry, i.e. with no additional Killing vector fields than the stationary and axial Killing vector fields.  The asymptotically flat vacuum region surrounding the rotating body can in that case be shown to have exactly these two Killing symmetries. It is plausible that all stationary, asymptotically flat spacetimes which are vacuum near infinity, are axisymmetric. \mnote{is this correct?}

%%%%%%%%%%%%%%%%%%%%%%%%%%%%%%%%%%%%%%%%%%%%%%%%%%%%%%%%%%%%%%%%%%%%%%%%%%%%%

\section{Dynamics of elastic bodies in general relativity} \label{sec:dynelast}
We write the Einstein-elastic system, cf. \eqref{eq:einstein-system}, in the form  
\begin{align*} 
R_{\mu\nu} &= 8 \pi G (T_{\mu\nu} - \half T g_{\mu\nu}) \chi_{f^{-1}(\Bo)}  \\ 
\nabla^\mu T_{\mu\nu} &= 0 \quad \text{ in $f^{-1}(\Bo)$ } \\ 
T_{\mu}{}^\nu \norm_\mu \big{|}_{\partial f^{-1}(\Bo)} &= 0
\end{align*} 
%Tension field: 
%$V^\alpha = g^{\mu\nu} \Gamma^\alpha_{\mu\nu} %- \hGamma^\alpha_{\mu\nu})$
In order to construct solutions to the Einstein equation it is convenient to work in wave coordinates gauge, 
\begin{equation}\label{eq:wavecoord}
g^{\mu\nu} \Gamma^\alpha_{\mu\nu} = 0
\end{equation} 
%$V^\alpha = 0$ 
A standard calculation, cf. \cite[Chapter 10.2]{wald:text} shows that with \eqref{eq:wavecoord} imposed, the Einstein equation takes the form becomes a  quasi-linear wave equation of the form  
$$
- \half \square_g g_{\mu\nu} + S_{\mu\nu} (g,\partial g) = 8\pi G(T_{\mu\nu} 
- \half T   g_{\mu\nu}) 
$$
where $\square_g = \nabla^\alpha \nabla_\alpha$ is the scalar d’Alembertian and $S_{\mu\nu}$ is is an expression which is quadratic in derivatives of $g_{\mu\nu}$.  Assuming suitable constitutive relations for the elastic material, the Einstein-elastic system now becomes a quasi-linear hyperbolic system, and one can proceed to construct solutions along standard lines. 

A serious obstacle however is the fact that the matter density has a jump at the surface of the body. This means that using standard techniques it appears difficult to prove local well-posedness for this system, even using sophisticated harmonic analysis techniques, as appears in the proof of the $L^2$ curvature conjecture. In a joint paper with Oliynyk  \cite{andersson:oliynyk:MR3150755} we have given a proof of local existence for solutions of quasi-linear systems with the appropriate discontinuity in the source term. There we have also given an outline of the application of the results of that paper to the Einstein-elastic system \cite[\S 5]{andersson:oliynyk:MR3150755}. Details will appear in a joint paper with Oliynyk and Schmidt \cite{AOS:dynelast}.

An important aspect of the problem can be seen by considering the following model problem. In $\Re^{n,1}$ with coordinates $(x^\alpha) = (t,x^i)$, 
let $\square = - \partial_t^2 + \Delta$ and consider the Cauchy 
problem
\begin{subequations}\label{eq:cauchy} 
\begin{align} 
\square u &= F(t,x,u, \partial u) \chi_\Omega, \\
u(0) &= u^0, \quad  \partial_t u(0) = u^1 .
\end{align} 
\end{subequations} 
Let $u_\ell = \partial_t^\ell u, \quad F_\ell = \partial_t^\ell F$ and let $s$ be a given, sufficiently large integer,  
\newcommand{\HH}{\mathcal H}
and let the spaces $\HH^{s}$ be defined by 
$$
\HH^{s} = 
\left \{ \begin{array}{ll} H^s(\Re^n), & s = 0, 1, \\ 
H^s(\Re^n) \cap H^s(\Omega) \cap H^s(\Re^n \setminus \Omega), & s \geq 2 .
\end{array} \right. 
$$
Suppose we are given data satisfying the \emph{compatibility conditions} 
\begin{equation}\label{eq:compatibility}
u_\ell(0) \in \HH^{s+1-\ell} ,
\end{equation} 
and assume that $F_\ell(\cdot,t) \in H^{s-\ell}(\Omega)$, for $0 \leq \ell \leq s$. 
Time differentiating the equation yields 
\begin{equation}\label{eq:uj}
\square u_j = F_j \chi_\Omega, \quad j = 0, \dots, s  .
\end{equation}
A standard energy estimate shows that $u_s, u_{s+1}$ are bounded in $H^1$ and $L^2$, respectively. 
One gets improved regularity for lower time derivatives by an induction argument. 
%Time differentiating the equation yields 
%From equation \eqref{eq:uj} for $j=\ell+1$ we have 
%\begin{equation}\label{eq:timediffeq}
%\Delta u_{s-1-\ell} = F_{s-1-\ell} \chi_\Omega + u_{s+1-\ell} , \quad \text{ $0 \leq \ell \leq s-1$.}
%\end{equation}
From \eqref{eq:uj} for $j=s-1$, we have 
\begin{align*} 
\Delta u_{s-1}  &= F_{s-1} \chi_\Omega + \partial_t^2 u_{s-1} \\
&= F_{s-1} \chi_\Omega + u_{s+1} .
\end{align*} 
The potential theory results mentioned in section \ref{sec:classelast} imply that $u_{s-1} \in \HH^2$. 
Suppose now we have for $\ell \geq 1$ an estimate for $u_{s-\ell}$ in $\HH^{\ell+1}$ in terms of the initial data and the bound on $F_{s-\ell}$ in $H^\ell(\Omega)$. 
Then we have from equation \eqref{eq:uj} for $j=s-1-\ell$, 
$$
\Delta u_{s-1-\ell} = F_{s-1-\ell} \chi_\Omega + u_{s+1-\ell} \in \HH^{\ell} .
$$
and the potential theory results we can now be used together with the assumptions on the initial data and $F$, to give an estimate for $u_{s-1-\ell}$ in $\HH^{\ell+2}$. 
Induction with $\ell=1$ as base yields an estimate for $u$ in $\HH^{s+1}$. 

\begin{figure} 
\centering
\psfrag{MAt0}{$\{t=0\}$}
\includegraphics[width=1.5in]{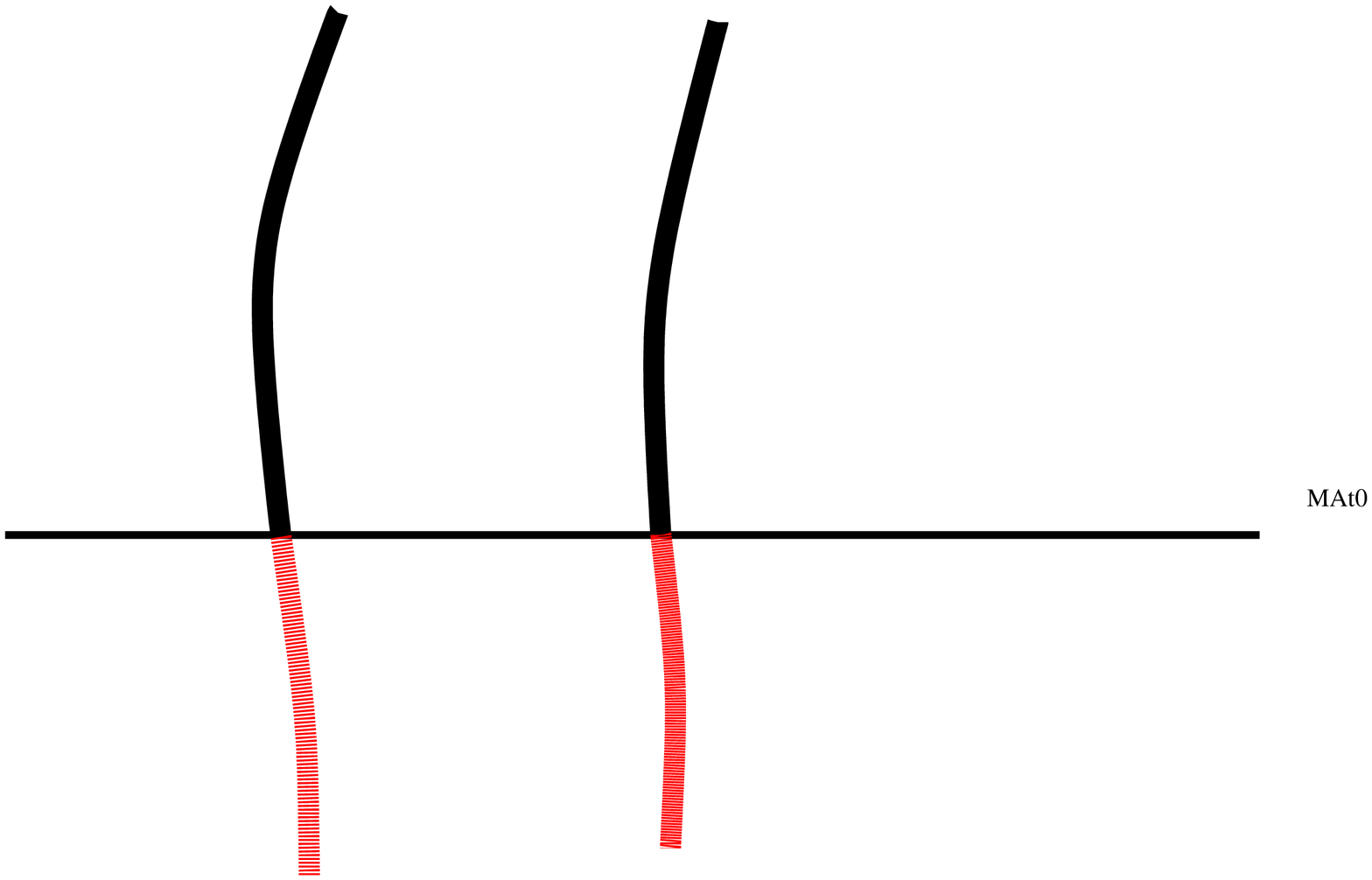} 
\caption{}
\label{fig:past2}
\end{figure}

An argument similar to the above forms an important part in the proofs of local well-posedness in the papers \cite{andersson:oliynyk:MR3150755,AOS:dynelast} mentioned above. 
The compatibiliary conditions \eqref{eq:compatibility} on initial data can be interpreted as implying that the body (or in the model problem, the source) existed and was regular in the past of the initial Cauchy surface, i.e. one must have the situation illustrated in figure \ref{fig:past2}. 

%%%%%%%%%%%%%%%%%%%%%%%%
\newcommand{\prd}{Phys. Rev. D}
%\bibliographystyle{abbrv}
%\bibliography{badhonnef}

\end{document}